\documentclass{SCNarticle} 
\usepackage{graphicx}
\usepackage{hyperref}
\usepackage{authblk}

\begin{document}


\title{A unified picture of phonon anomalies in crystals and glasses}
\author[1]{Alessio Zaccone}
\affil[1]{Department of Physics "A. Pontremoli", University of Milan, via Celoria 16, 20133 Milano, Italy}

\maketitle

\noindent
Phonons, the quantized vibrations of atoms in solids, usually follow the Debye law: the vibrational density of states grows as the square of the vibration frequency. Yet real materials often deviate from this simple picture. Crystals exhibit sharp \emph{van Hove singularities} (VHS), while glasses show the smoother but puzzling \emph{boson peak} (BP). For decades these anomalies were studied separately. A key question has remained: could they share a universal origin?  

A recent work by Ding \emph{et al.}~\cite{Ding2025} in \emph{Nature Physics} offers a compelling answer. Their continuum model demonstrates that both VHS and BP can arise from the same physical mechanism: the competition between phonon propagation and phonon damping, accompanied by vibrational softening. The theory predicts a phase diagram that distinguishes materials showing VHS, BP, or both, and successfully accounts for specific-heat data across 143 different real solids.  


The idea that damping underlies vibrational anomalies was proposed earlier within the Baggioli-Zaccone (BZ) theory~\cite{BZ2019,BZ2020}. BZ argued that phonons, even in perfect crystals, are damped due to anharmonicity. As damping increases, propagating modes cross over into diffusive-type modes, leaving a boson-peak signature. The message is clear: boson peaks are not confined to disordered solids, but represent a universal competition between phonon propagation and phonon damping. Wei-Hua Wang and co-workers successfully used the BZ model to describe and explain the boson peak in the heat capacity of many different materials~\cite{Wei-Hua}. Importantly, this theory does not require any \emph{ad hoc} assumption on random spatial distributions of elastic moduli, as in older heterogeneous elasticity theories of the boson peak, which are untenable in crystals.

Ding \emph{et al.} extend the BZ concept in a crucial way. Rather than assuming a quadratic damping law, they derive a frequency- and wavevector-dependent damping function from scattering by local resonators. Importantly, damping is coupled to phonon softening: as damping grows, the effective phonon frequency is reduced, further amplifying deviations from the Debye scaling. Their model naturally explains why some materials exhibit a single VHS, others a BP, and still others a coexistence of both.  


In glasses, low-frequency phonons in a certain range are known to obey the Rayleigh law, where attenuation grows as the fourth power of frequency. Microscopically, as shown rigorously in~\cite{BaggioliZaccone2022JPCM} this originates from nonaffine atomic motions: atoms in disordered environments fail to move affinely, generating random forces that scatter phonons. A theoretical framework of nonaffine elasticity elucidates how this mechanism reduces elastic moduli~\cite{Zaccone2011,ZacconeBook} and enhances wave damping~\cite{BaggioliZaccone2022JPCM}.  

The resonant-damping model of Ding \emph{et al.} incorporates this low-\(q\) Rayleigh regime while extending to higher-\(q\) resonant scattering. In this way, both crystals and glasses can be understood within a unified framework based on the underlying phonon propagating and damping relations, as shown in Fig. 1.  


Recent experiments strongly support this picture. Wang \emph{et al.}~\cite{Wang2025} observed quartic attenuation scaling in silica glass, consistent with Rayleigh scattering, together with negative dispersion in acoustic velocity—direct evidence of damping-induced softening. Boson-like anomalies have also been reported in strain-glass systems~\cite{Ren2021}, while boson peaks in both glassy and complex crystalline systems (perosvkites, intermetallics etc) have been linked to anharmonic damping by several authors~\cite{Zhao2025PRL,Cheng2025}, in agreement with the BZ proposal. 

Altogether, these findings reinforce the emerging consensus: boson peaks and van Hove singularities are not completely distinct phenomena, but two different aspects of the same underlying phonon physics.  


Hence, the resonant-damping framework represents a major advance toward a unified theory of vibrational anomalies. Challenges remain, notably mapping its effective parameters (scatterer size and mean free path) onto real microstructural features such as defects or clusters. Direct measurements of phonon damping functions in simulations and inelastic scattering experiments will be crucial tests. Extending the theory to finite temperatures and transport properties, such as thermal conductivity, would further broaden the impact of the framework.  

By combining elegant conceptual ideas with quantitative predictive power, the work of Ding \emph{et al.}~\cite{Ding2025} brings us closer to resolving a long-standing puzzle of condensed matter physics: the true origin of vibrational anomalies in solids.  

\begin{figure}[t]
\centering
\includegraphics[width=0.95\textwidth]{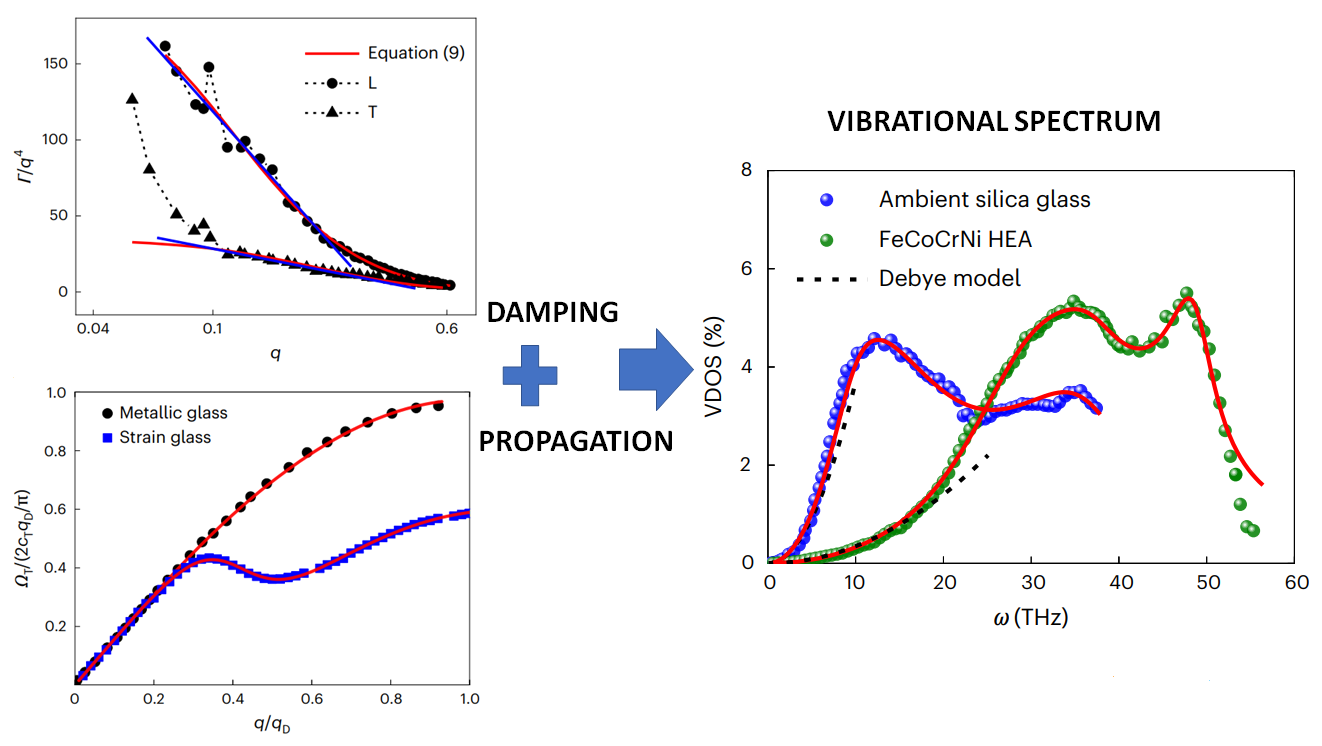}
\caption{(Color online) Unified picture of vibrational anomalies. 
Left: phonon damping functions derived from resonant scattering reproduce both Rayleigh-like (\(\Gamma \propto q^4\)) behavior at low wavevectors and resonant softening at higher \(q\), consistent with data from glasses and crystals. 
Right: the resulting vibrational density of states (VDOS) explains boson peaks in glasses (silica, blue) and van Hove-like features in high-entropy alloys (FeCoCrNi HEA, green), in comparison with the Debye model.  
Adapted from Ding \emph{et al.}~\cite{Ding2025}.}
\end{figure}



\end{document}